\documentclass[aip,apl,amsmath,amssymb,reprint]{revtex4-1}
\usepackage{color, graphicx}
\usepackage[version=3]{mhchem} 
\usepackage{subfig}
\usepackage{dcolumn} 
\usepackage{bm} 
\usepackage[colorlinks=true,linkcolor=blue,citecolor=blue,urlcolor=blue]{hyperref}

\begin{document}

\title{Role of electron-phonon coupling and thermal expansion on band gaps, carrier mobility, and interfacial offsets in kesterite thin-film solar cells}

\author{Bartomeu Monserrat}
\affiliation{Department of Physics and Astronomy, Rutgers University, Piscataway, New Jersey 08854-8019, USA}
\affiliation{Cavendish Laboratory, University of Cambridge, J. J. Thomson Avenue, Cambridge CB3 0HE, United Kingdom}

\author{Ji-Sang Park}
\author{Sunghyun Kim}
\affiliation{Department of Materials, Imperial College London, Exhibition Road, London SW7 2AZ, UK}

\author{Aron Walsh}
\affiliation{Department of Materials, Imperial College London, Exhibition Road, London SW7 2AZ, UK}
\affiliation{Department of Materials Science and Engineering, Yonsei University, Seoul 03722, Korea}

\date{\today}

\begin{abstract}
The efficiencies of solar cells based on kesterite \ce{Cu2ZnSnS4} (CZTS) and \ce{Cu2ZnSnSe4} (CZTSe) are limited by a low open-circuit voltage due to high rates of non-radiative electron-hole recombination.  To probe the origin of this bottleneck, we calculate the band offset of CZTS(Se) with CdS, confirming a weak spike of $0.1$\,eV for CZTS/wurtzite-CdS and a strong spike of $0.4$\,eV for CZTSe/wurtzite-CdS. We also consider the effects of temperature on the band alignment, finding that increasing temperature significantly enhances the spike-type offset. We further resolve an outstanding discrepancy between measured and calculated phonon frequencies for the kesterites, and use these to estimate the upper limit of electron and hole mobilities based on optic phonon Fr\"ohlich scattering, which uncovers an intrinsic asymmetry with faster (minority carrier) electron mobility. 
\end{abstract}


\maketitle 

The kesterite semiconductors Cu$_2$ZnSnS$_4$ (CZTS) and Cu$_2$ZnSnSe$_4$ (CZTSe) provide a promising route towards next-generation photovoltaics,\cite{walsh2012review,mitzi2013review,walsh2017review} with certified solar conversion efficiencies above $12$\%.\cite{mitzi2014,kim2016} Attractive features of these materials include the presence of only earth-abundant non-toxic elements and the high tunability of their optical band gap\cite{wei2009,zhao2011optical} in the range relevant for single-junction solar cell applications.\cite{shockley_queisser1961} 

The biggest challenge facing CZTS(Se) solar cells is a large open-circuit voltage deficit, which becomes clear when the device parameters are compared to other thin-film photovoltaic technologies.\cite{mitzi2014} 
The low p-type doping efficiency, short electron minority carrier lifetime, and interface recombination have been considered as potential origins of the problem.\cite{kim2017strategies} Device simulations show that interface recombination significantly lowers the open-circuit voltage when the n-type electron extraction window layer has a lower conduction band minimum (CBM) than the absorber layer (cliff-type offset).\cite{gloeckler2005efficiency} 
As such, the conduction band offset should be carefully controlled to maximize efficiency by making a preferred spike-type band offset. However, there is no consensus on the nature of the conduction band offset of the CZTS(Se)/CdS interface.\cite{crovetto2017,kim2017strategies}

In this work, we report a first-principles investigation of the temperature dependence of the band edges of CZTS, CZTSe, and CdS, which allows us to assess the band alignment between CZTS(Se) and CdS. We find a weak spike-type band offset in CZTS/wurtzite-CdS, which increases in size with increasing temperature to a value of $0.17$\,eV at $300$\,K. We also find that the band alignment varies with polymorph of CdS and the choice of chalcogen in the kesterite. Additionally, we resolve an outstanding discrepancy between the calculated and measured vibrational frequencies of the kesterites by including exact exchange via a hybrid density functional, and the revised phonon dispersion is used to compute the polar-scattering limited carrier mobilities.

\textit{Computational Methods.}
We have performed calculations on CZTS and CZTSe in the kesterite structure, and on CdS in the wurzite (\textit{wz}-CdS) and zinc-blende (\textit{zb}-CdS) polymorphs, using density functional theory (DFT) in the projector augmented-wave method~\cite{paw_original,paw_us_relation} as implemented in {\sc vasp}.~\cite{vasp1,vasp2,vasp3,vasp4} 
We have used the Perdew-Burke-Ernzerhof semilocal functional (PBE) \cite{pbe_functional}, the PBE functional for solids (PBEsol),\cite{pbesol_functional} and the Heyd-Scuseria-Ernzerhof hybrid functional (HSE06). \cite{hse03_functional,hse06_functional,hse06_functional_erratum} 
The electronic structure has been calculated with an energy cut-off of $550$\,eV and an electronic Brillouin zone (BZ) sampling grid of size $4\times4\times4$ for the kesterites, and $8\times8\times8$ for CdS, and commensurate grids for the supercell calculations.  

The temperature dependence of the electronic band structure is caused by thermal expansion and electron-phonon coupling. 
The starting point for evaluating both contributions is the harmonic approximation to lattice dynamics, which we have calculated using the finite differences method \cite{phonon_finite_displacement} in conjunction with non-diagonal supercells.\cite{non_diagonal,monserrat2018review} 
We have used a coarse $\mathbf{q}$-point grid of $4\times4\times4$ points to sample the vibrational BZ, and Fourier interpolation to a fine grid to obtain accurate vibrational free energies and densities of states. 
Thermal expansion is calculated within the quasiharmonic approximation.\cite{dove_lattice_dynamics_book} 
We have calculated the electron-phonon coupling contribution using a single-phonon theory first proposed by Allen and Heine \cite{allen_heine_ep} in which the finite temperature value of an electronic eigenvalue $(\mathbf{k},n)$ is:\cite{giustino_elph_diamond,monserrat_elph_diamond_silicon,gonze_marini_elph}
\begin{equation}
\epsilon_{\mathbf{k}n}(T)=\epsilon_{\mathbf{k}n}(\mathbf{0})+\sum_{\mathbf{q},\nu}\frac{1}{2\omega_{\mathbf{q}\nu}}\frac{\partial^2\epsilon_{\mathbf{k}n}}{du_{\mathbf{q}\nu}^2}\left[\frac{1}{2}+n_{\mathrm{B}}(\omega_{\mathbf{q}\nu},T)\right],
\end{equation}
where $\epsilon_{\mathbf{k}n}(\mathbf{0})$ is the static lattice eigenvalue, $\omega_{\mathbf{q}\nu}$ is the harmonic frequency of a phonon of wave vector $\mathbf{q}$ and branch $\nu$, $u_{\mathbf{q}\nu}$ is the normal mode amplitude of that phonon, and $n_{\mathrm{B}}$ is the Bose-Einstein factor. We have also calculated the finite temperature eigenvalues using Monte Carlo integration~\cite{giustino_nat_comm,molec_crystals_elph} to confirm that multi-phonon terms are negligible in the studied systems. Convergence details of the calculations are provided in the Supplementary Material, and a recent review of the methods used can be found in Ref.~\onlinecite{monserrat2018review}.

We have determined band offsets for the CZTS(Se)/CdS interface using the procedure proposed by Li \textit{et al.}\cite{li2009revised}
We have constructed (001) \textit{zb}-CdS/(001) CZTS(Se) interfaces with $7$ double-layers each.
The lattice constant of the supercell within the interface plane has been fixed to that of CZTS(Se). 
The CZTS(Se) layer at the (001) \textit{zb}-CdS/(001) CZTS(Se) interface can contain either Cu and Zn atoms or Cu and Sn atoms.
Owing to the distinct oxidation states of Zn(II) and Sn(IV), a different dipole field  is formed at each interface, 
and the potential difference has been averaged to compensate them.
The valence band maximum (VBM) of \textit{wz}-CdS has been estimated using the valence band offset between \textit{wz}-CdS and \textit{zb}-CdS of $46$\,meV.\cite{wei2000structure}
The corresponding conduction band offset, calculated using HSE06, is $115$\,meV.

\begin{figure}
\begin{center}
\resizebox{0.8\columnwidth}{!}{\includegraphics*{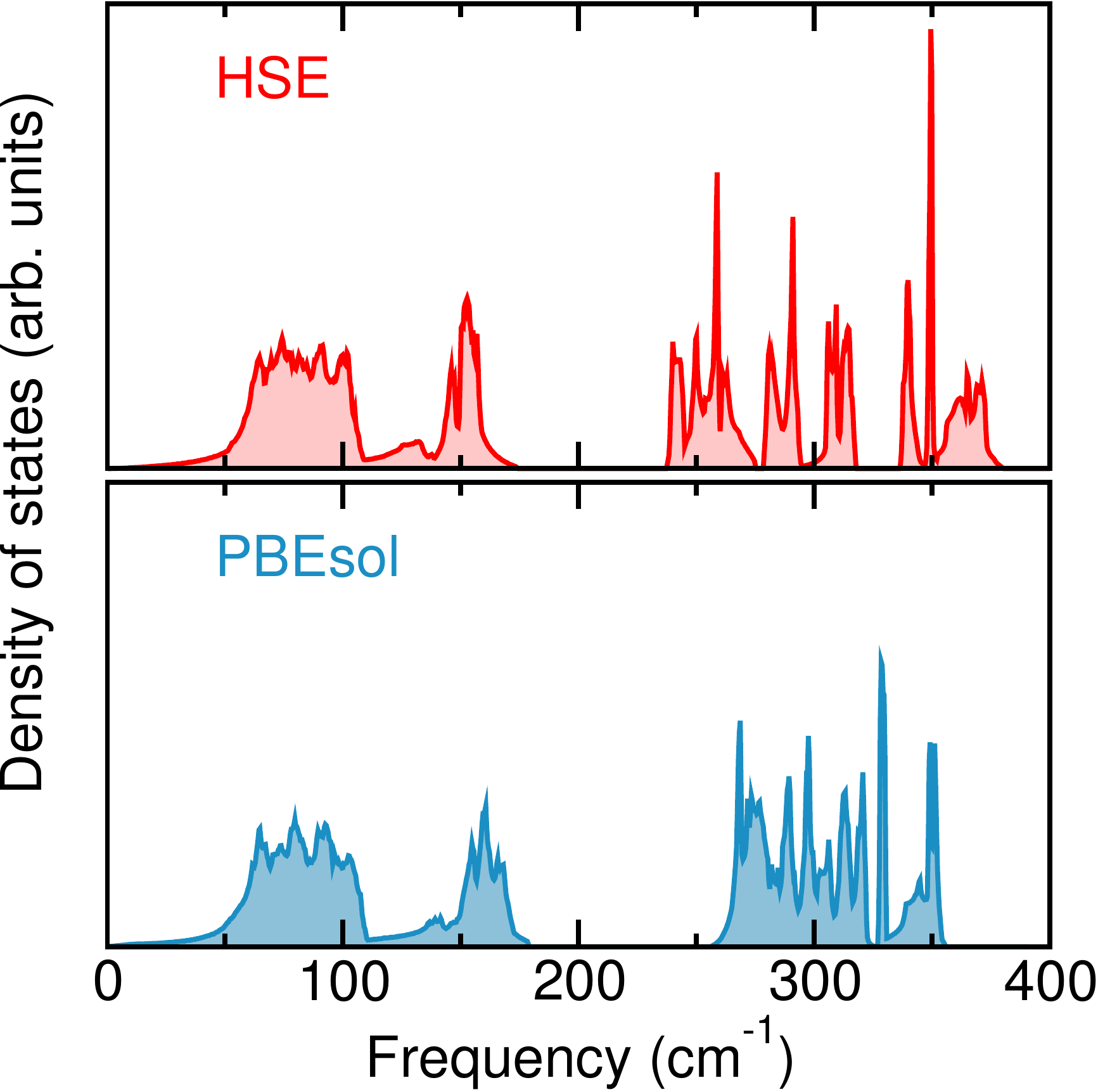}}
\caption{\label{f1} Vibrational density of states of CZTS calculated using the HSE06 (top) and PBEsol (bottom) functionals. The phonon frequency range at the HSE06 level extends to higher frequencies, in good agreement with experimental measurements. A similar picture emerges for CZTSe (see Supplementary Material).} 
\end{center}
\end{figure}

\textit{Lattice Dynamics.}
Khare and co-workers\cite{aydil_2012} suggested that the Raman active vibrational modes of CZTS and CZTSe could be used to distinguish between different polymorphs (e.g. kesterite and stannite). 
In this context, the lattice dynamics of CZTS and CZTSe have been extensively studied using semilocal DFT, \cite{aydil_2012,cztse_dft_postnikov_2010,czts_dft_cagin_2011,jackson_2014,skelton_2015,gonze_2016} and our PBEsol results for CZTS, shown in Fig.~\ref{f1} (bottom), are in good agreement with earlier reports. The phonon density of states exhibits low energy phonons up to frequencies of about $180$\,cm$^{-1}$ dominated by copper, zinc, and tin nuclear motion, and high energy phonons in the range $260$-$360$\,cm$^{-1}$ dominated by sulfur motion. 

The available semilocal DFT calculations fail to quantitatively describe the experimentally observed vibrational frequencies for CZTS, with the high-energy modes appearing about $15$\,cm$^{-1}$ softer in energy compared to experiment.\cite{czts_raman_exp_2014} This discrepancy has been attributed to the overestimation of the polarisability of S atoms in semilocal DFT\cite{skelton_2015} and to the disorder\cite{skelton_2015} and polymorphism\cite{gonze_2016} in the experimental samples. Unfortunately, the frequency differences between polymorphs are comparable to the theory-experiment discrepancy,\cite{gonze_2016} precluding the use of Raman spectra to determine the polymorphism of experimental samples. We resolve this discrepancy by performing lattice dynamics using the hybrid HSE06 functional,\cite{hse03_functional,hse06_functional,hse06_functional_erratum} which provides more accurate polarisabilities than semilocal DFT.\cite{amos_polarisability} Our results in Fig.~\ref{f1} (top) show that the inclusion of nonlocal exhange shifts the high-energy modes by about $15$\,cm$^{-1}$, and the resulting frequencies are in excellent agreement with experiment.\cite{czts_raman_exp_2014} We reach similar conclusions for CZTSe (see Supplementary Material).

\begin{table}[b]
  \setlength{\tabcolsep}{4pt} 
  \caption{\label{t1}  Fr\"ohlich electron-phonon interaction parameter ($\alpha$) for \ce{Cu2ZnSnS4} (CZTS) and \ce{Cu2ZnSnSe4} (CZTSe) along with values for the electron ($e$) and hole ($h$) mobilities ($\mu$) at $300$\,K, and high-frequency ($\epsilon_\infty$) and static dielectric constants ($\epsilon_0$). $m^*$ represents the harmonic mean of the carrier effective mass in units of $m_e$.}
  \label{tbl:properties}
\begin{tabular}{lcccccc}
    \hline    \hline
    Material & Carrier & $m^*$ & $\epsilon_\infty$ & $\epsilon_0$ & $\alpha$ & $\mu$ (cm$^2$ V$^{-1}$ s$^{-1}$) \\    \hline
    CZTS  & $e$        & 0.18  & 7.1               &  9.9         & 0.35 &  544 \\
    CZTS  & $h$        & 0.40  &                   &              & 0.53 & 159 \\
    CZTSe & $e$        & 0.10  & 8.1               & 11.4         & 0.29 & 1291 \\
    CZTSe & $h$        & 0.23  &                   &              & 0.44 & 361 \\
    \hline
	\hline
  \end{tabular}
\end{table}

\textit{Carrier Mobility.}
The mobility of carriers in semiconductors is determined by a range of scattering processes including impurity scattering, carrier-carrier scattering, and electron-phonon scattering. 
However, the room temperature mobility in polar semiconductors is often limited by scattering with optic-phonon modes, which can be described in the  Fr\"ohlich large polaron model.\cite{frohlich1939mean} 
Taking a recent implementation\cite{frost2017calculating} of the Feynman path-integral approach,\cite{feynman1955slow} we have used our calculated phonon frequencies to estimate the Fr\"ohlich interaction $\alpha$, in addition to an upper limit for the room temperature mobility.  
The effective mass of electrons is smaller than holes, 
which results in a pronounced asymmetry in the mobility values (Table \ref{t1})
with a ratio greater than 3:1 for both CZTS and CZTSe.
%
%
These compare to recent Hall measurements of $\mu_e = 10$ cm$^2$ V$^{-1}$ s$^{-1}$ and $\mu_h = 1$ cm$^2$ V$^{-1}$ s$^{-1}$ for high-quality CZTSSe.\cite{gunawan2018carrier}  
Both measured mobilities are small, which implies additional scattering processes are active, but the stronger asymmetry of 10:1 between 
electron and hole mobilities suggests that a carrier selective process (i.e. a deep hole trap) is severely limiting hole transport to well below its intrinsic limits.

\begin{figure}
\begin{center}
\resizebox{0.8\columnwidth}{!}{\includegraphics*{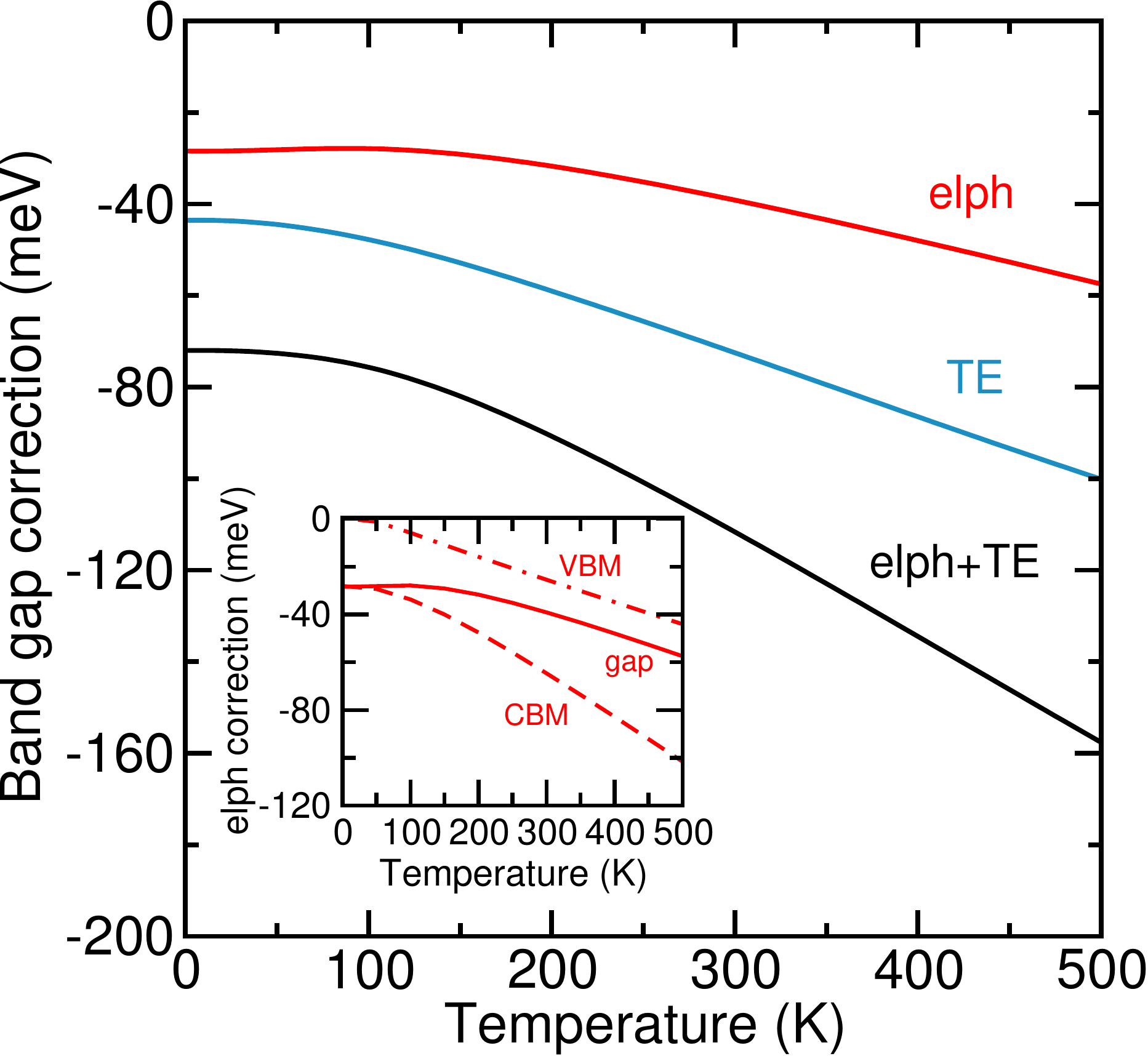}}
\caption{\label{f2} Temperature dependence of the band gap of CZTS arising from electron-phonon coupling only (elph), thermal expansion only (TE), and the total dependence (elph+TE). Inset: temperature dependence of the band edges arising from electron-phonon coupling.}
\end{center}
\end{figure}

\textit{Band Gap Renormalization.}
The calculated temperature dependence of the band gap of CZTS is shown in Fig.~\ref{f2}. The thermal expansion (TE) and electron-phonon coupling (elph) contributions are considered separately, and thermal expansion dominates. However, if the temperature dependences of the individual band edges are considered, then the electron-phonon coupling contribution is larger (see inset in Fig.~\ref{f2}). This is because both VBM and CBM decrease in value with increasing temperature, partially cancelling each other when calculating the temperature dependence of the band gap. 

To investigate the microscopic origin of the temperature dependence reported in Fig.~\ref{f2} we have identified the phonon modes that provide the dominant contribution to the electron-phonon coupling renormalisation (see Supplementary Material). These are the high-energy modes which correspond to vibrations of the sulfur atoms, suggesting that cation disorder in Cu and Zn will not significantly modify the strength of electron-phonon coupling. We have also determined that the band gap change is largest for thermal expansion along the $a$ axis of the kesterite structure.

We have also calculated the temperature dependence of the crystal field splitting of the valence bands $\Gamma_{\mathrm{5v}}-\Gamma_{\mathrm{4v}}$, and found that it decreases from $-54$\,meV at $0$\,K to $-18$\,meV at $500$\,K (see Supplementary Material). 

We finally note that the electron-phonon coupling calculations have been performed using the hybrid HSE06 functional because the corresponding PBEsol calculations were ill-behaved due to unrealistically small band gaps of $0.12$\,eV for CZTS (compared to $1.47$\,eV using HSE06) and metallic behaviour for CZTSe (compared to $0.90$\,eV using HSE06). The HSE06 band gaps are in good agreement with experimental reports for both compounds.\cite{yuan2016}

\begin{table}[b]
  \setlength{\tabcolsep}{6pt} 
  \caption{\label{t2} Calculated static-lattice band offsets of CZTS(Se)/CdS. Positive band offset for an interface A/B indicates that A has lower band than B.}
  \label{tbl:properties}
\begin{tabular}{lcc}
    \hline    \hline
    Interface & CBO (eV) & VBO (eV) \\    \hline
    CZTS/\textit{wz}-CdS & 0.10 & -0.70\\
    CZTS/\textit{zb}-CdS & -0.01  &-0.75\\
    CZTSe/\textit{wz}-CdS & 0.39  & -0.99\\
    CZTSe/\textit{zb}-CdS & 0.28  & -1.04\\
    \hline
	\hline
  \end{tabular}
\end{table}

\textit{Band Offsets.}
Conduction band offsets (CBO) and valence band offsets (VBO) of the kesterites and two polymorphs of CdS are shown in Table~\ref{t2}.
Our calculations confirm that the conduction band of \textit{wz}-CdS is $0.1$\,eV higher in energy than the CBM of CZTS, making the interface a weak spike. 
This result is close to the average of the reported calculation values.\cite{chen2010intrinsic,nagoya2011first,dong2014experimental} 
Although a broad distribution of the CBO values have been reported in experiments, 
the CBO tends to increase with the device efficiency,\cite{crovetto2017} 
and our value is closer to those ($-0.13\sim0.10$) measured from solar cells with high efficiencies ($\ge$ 7\%).\cite{tajima2013direct,terada2015characterization,kataoka2016band}
The lower CBO values observed in CZTS cells with lower efficiencies can be caused by Fermi level pinning as discussed by Crovetto \textit{et al}.\cite{crovetto2017}

On the other hand, CZTSe has a lower conduction band and therefore a larger CBO of about $0.39$\,eV, in reasonable agreement with previous experimental and computational values.\cite{chen2011compositional,palsgaard2016semiconductor}
If \textit{zb}-CdS is locally formed at the interface, then the CBO could be reduced because of the lower conduction band of the zinc-blende phase. 
An additional possibility not considered here is intermixing between Zn and Cd, which could also be the origin of some the observed variations. \ce{Zn_{Cd}} in CdS is expected to raise the conduction band, while \ce{Cd_{Zn}} in CZTS would be expected to lower the conduction band, so intermixing can effectively change the nature of the junction. Finally, some of us have recently reported that stacking faults and antisite domain boundaries also modify the location of the conduction band.\cite{park2018}

\begin{figure}
\begin{center}
\resizebox{0.8\columnwidth}{!}{\includegraphics*{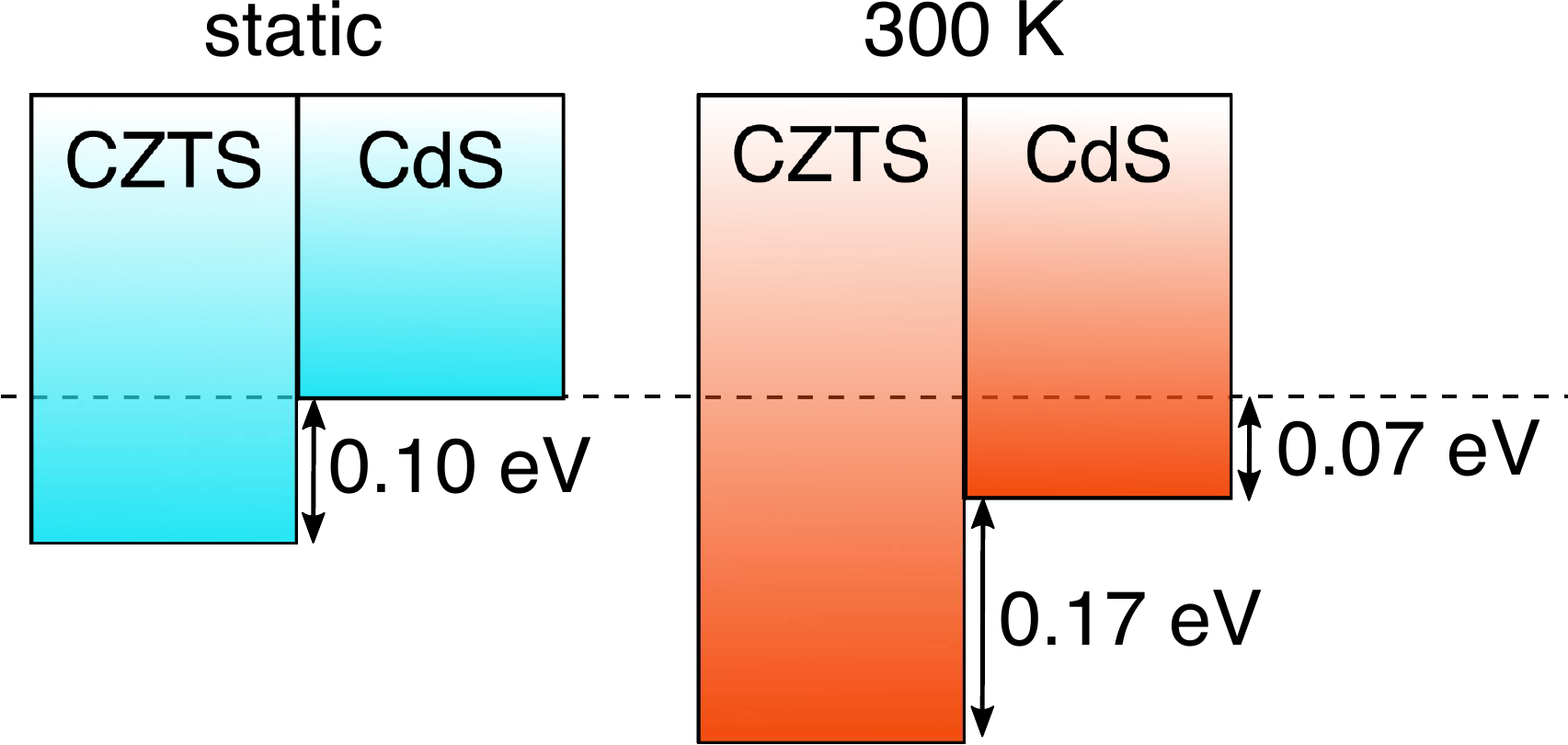}}
\caption{\label{f3} Calculated conduction band offset of \ce{Cu2ZnSnS4}/\textit{wz}-CdS at the static lattice level and at $300$\,K.}
\end{center}
\end{figure}

We use the thermal expansion and electron-phonon coupling results reported above to calculate the finite temperature CBO of CZTS/\textit{wz}-CdS. The thermal expansion contribution to the band gap change has been fully assigned to the CBM, as earlier work suggests that changes in volume mostly affect the CBM in CdS.\cite{li2006ab} The CBO of CZTS/\textit{wz}-CdS at $300$\,K is reported in Fig.~\ref{f3}, where we see that the static lattice CBO of $0.10$\,eV increases to $0.17$\,eV at $300$\,K. Our results imply that increasing temperature reinforces the spike-type offset in CZTS/\textit{wz}-CdS.  


In conclusion, we have used a first-principles lattice dynamics approach to account for the finite-temperature properties of kesterite semiconductors. 
We have resolved a discrepancy in the frequency range of the optic phonon branch using a hybrid exchange-correlation functional, and further used the phonon frequencies to predict an upper limit on the room temperature carrier mobilities. We have also confirmed that the natural band offset between CdS and CZTS is spike-like and is further enhanced (from $0.10$ to $0.17$ eV) by taking into account finite-temperature effects on the electronic structure. 

\section*{Supplementary Material}

See supplementary material for details of the electron-phonon coupling calculations for CZTS and CdS, the thermal expansion calculations, and the CZTSe calculations. 

\acknowledgments
We are grateful to the UK Materials and Molecular Modelling Hub for computational resources, which is partially funded by EPSRC (EP/P020194/1).
The research was supported by the Royal Society and the  EU Horizon2020 Framework (STARCELL, grant no. 720907).
B.M. thanks Robinson College, Cambridge, and the Cambridge Philosophical Society for a Henslow Research Fellowship.
J.P. thanks the Royal Society for a Shooter International Fellowship.

\textit{Data access statement:} The mobility calculations were performed using \textsc{PolaronMobility.jl} available from \url{https://github.com/jarvist/PolaronMobility.jl}. 

\bibliography{library}

\onecolumngrid
\clearpage
\begin{center}
\textbf{\large Supplementary Material for ``Role of electron-phonon coupling and thermal expansion on band gaps, carrier mobility, and interfacial offsets in kesterite thin-film solar cells''}
\end{center}
\setcounter{equation}{0}
\setcounter{figure}{0}
\setcounter{table}{0}
\setcounter{page}{1}
\makeatletter
\renewcommand{\theequation}{S\arabic{equation}}
\renewcommand{\thefigure}{S\arabic{figure}}
\renewcommand{\bibnumfmt}[1]{[S#1]}
\renewcommand{\citenumfont}[1]{S#1}

\section{Electron-phonon coupling calculations for C\lowercase{u}$_2$Z\lowercase{n}S\lowercase{n}S$_4$}

In this section we test the validity of the electron-phonon coupling calculations reported in the main text for Cu$_2$ZnSnS$_4$ (CZTS).

\subsection{Quadratic approximation}

In the main text we use the quadratic approximation to evaluate the electron-phonon coupling contribution to the temperature dependence of the band gap. The adiabatic finite temperature value of an electronic eigenvalue is given by
\begin{equation}
\epsilon_{\mathbf{k}n}(T)=\frac{1}{\mathcal{Z}}\sum_{\mathbf{s}}\langle\chi_{\mathbf{s}}(\mathbf{u})|\epsilon_{\mathbf{k}n}(\mathbf{u})|\chi_{\mathbf{s}}(\mathbf{u})\rangle e^{-E_{\mathbf{s}}/k_{\mathrm{B}}T}, \label{eq:elph}
\end{equation}
where $\mathcal{Z}=\sum_{\mathbf{s}}e^{-E_{\mathbf{s}}/k_{\mathrm{B}}T}$ is the partition function, $|\chi_{\mathbf{s}}(\mathbf{u})\rangle$ and $E_{\mathbf{s}}$ are the vibrational wavefunction and energy in state $\mathbf{s}$, $\mathbf{u}=\{u_{\mathbf{q}\nu}\}$ is a collective coordinate for all normal modes of vibration, $T$ is the temperature, and $k_{\mathrm{B}}$ is Boltzmann's constant. Assuming that the lattice dynamics are harmonic, we can evaluate Eq.~(\ref{eq:elph}) using Monte Carlo (MC) integration:
\begin{equation} 
\epsilon_{\mathbf{k}n}(T)\simeq\frac{1}{N}\sum_{i=1}^{N}\epsilon_{\mathbf{k}n}(\mathbf{u}_i),
\end{equation}
where the sampling points $\mathbf{u}_i$ are distributed according to the harmonic vibrational density at each temperature. This expression makes no assumption about the dependence of $\epsilon_{\mathbf{k}n}(\mathbf{u})$ on the atomic configuration.

Due to the computational expense of the MC calculation, arising mostly from the need to use large supercells to sample the vibrational Brillouin zone, it is useful to expand $\epsilon_{\mathbf{k}n}(\mathbf{u})$ about its equilibrium position and to truncate the expansion at second order. This approach leads to the quadratic approximation used in the main manuscript, that we repeat here for completeness:\cite{allen_heine_ep}
\begin{equation}
\epsilon_{\mathbf{k}n}(T)=\epsilon_{\mathbf{k}n}(\mathbf{0})+\sum_{\mathbf{q},\nu}\frac{1}{2\omega_{\mathbf{q}\nu}}\frac{\partial^2\epsilon_{\mathbf{k}n}}{du_{\mathbf{q}\nu}^2}\left[\frac{1}{2}+n_{\mathrm{B}}(\omega_{\mathbf{q}\nu},T)\right]. \label{eq:quad}
\end{equation}
The quadratic approximation dramatically reduces the computational expense of the calculations as we can now exploit nondiagonal supercells,~\cite{non_diagonal} but at the cost of neglecting multi-phonon terms in the electron-phonon interaction.

\begin{figure}
\begin{center}
\includegraphics[width=0.35\textwidth]{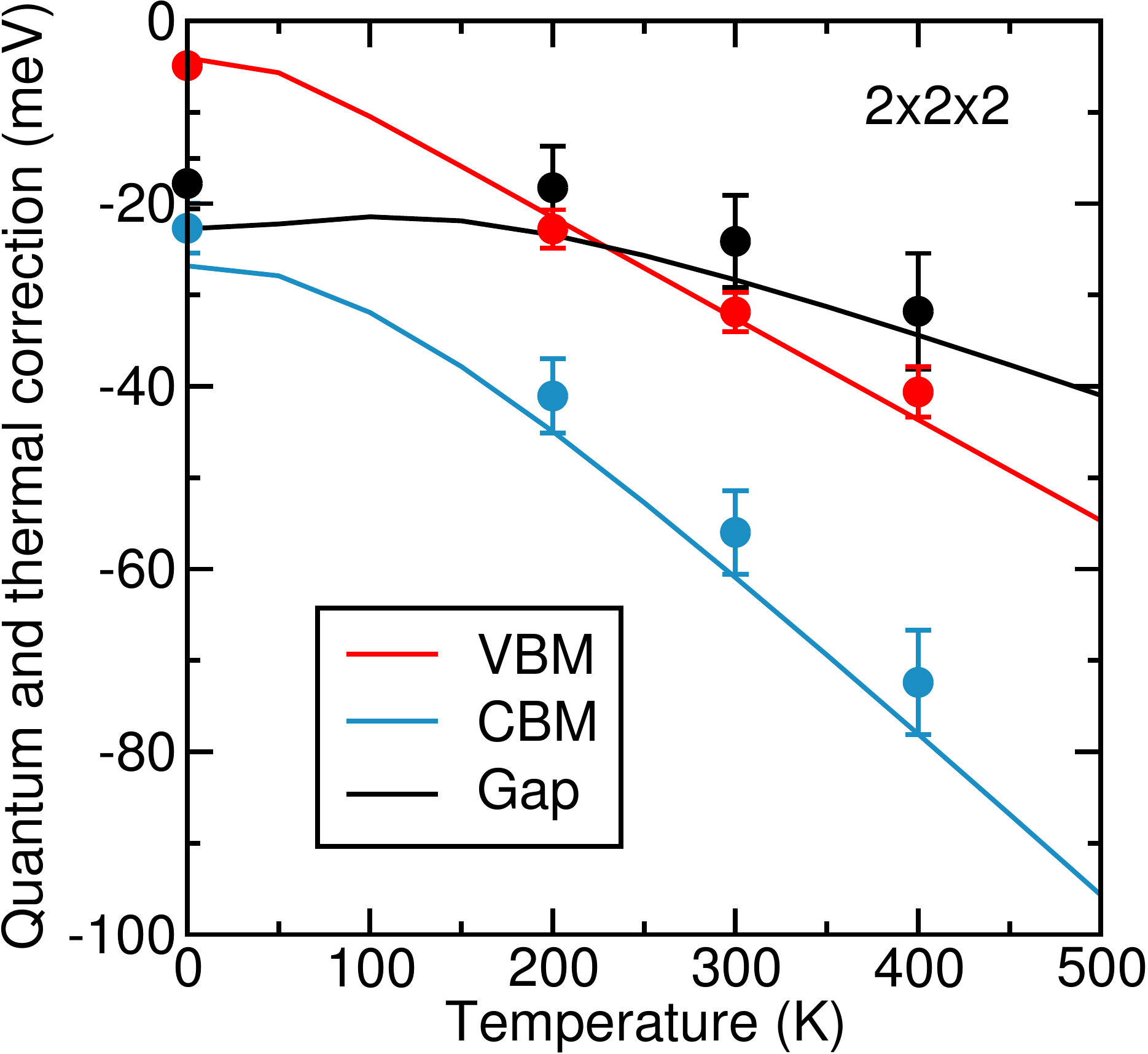}
\caption{{\footnotesize Vibrational correction (quantum and thermal) to the valence band maximum, conduction band minimum, and overall band gap of CZTS. The circles correspond to MC results, and the lines to the quadratic approximation. The calculations have been performed using the hybrid HSE06 functional\cite{hse03_functional,hse06_functional,hse06_functional_erratum} and sampling the vibrational Brillouin zone with a $2\times2\times2$ $\mathbf{q}$-point grid.}}
\label{fig:quad_vs_mc}
\end{center}
\end{figure}

We have evaluated the temperature dependence of the band gap of CZTS using both the MC and quadratic approaches for a $2\times2\times2$ $\mathbf{q}$-point grid of the vibrational Brillouin zone. The rest of numerical parameters of the calculation are the same as those reported in the main manuscript. A comparison between the two approaches is provided in Fig.~\ref{fig:quad_vs_mc}, where we report the temperature dependence of the valence band maximum (VBM), the conduction band minimum (CBM), and the overall band gap. The MC results are depicted with circles, and the quadratic approximation results with lines. Good agreement is observed between the two, confirming the validity of the quadratic approximation used in the main manuscript.

\subsection{Vibrational Brillouin zone sampling}

\begin{figure}
\begin{center}
\includegraphics[width=0.35\textwidth]{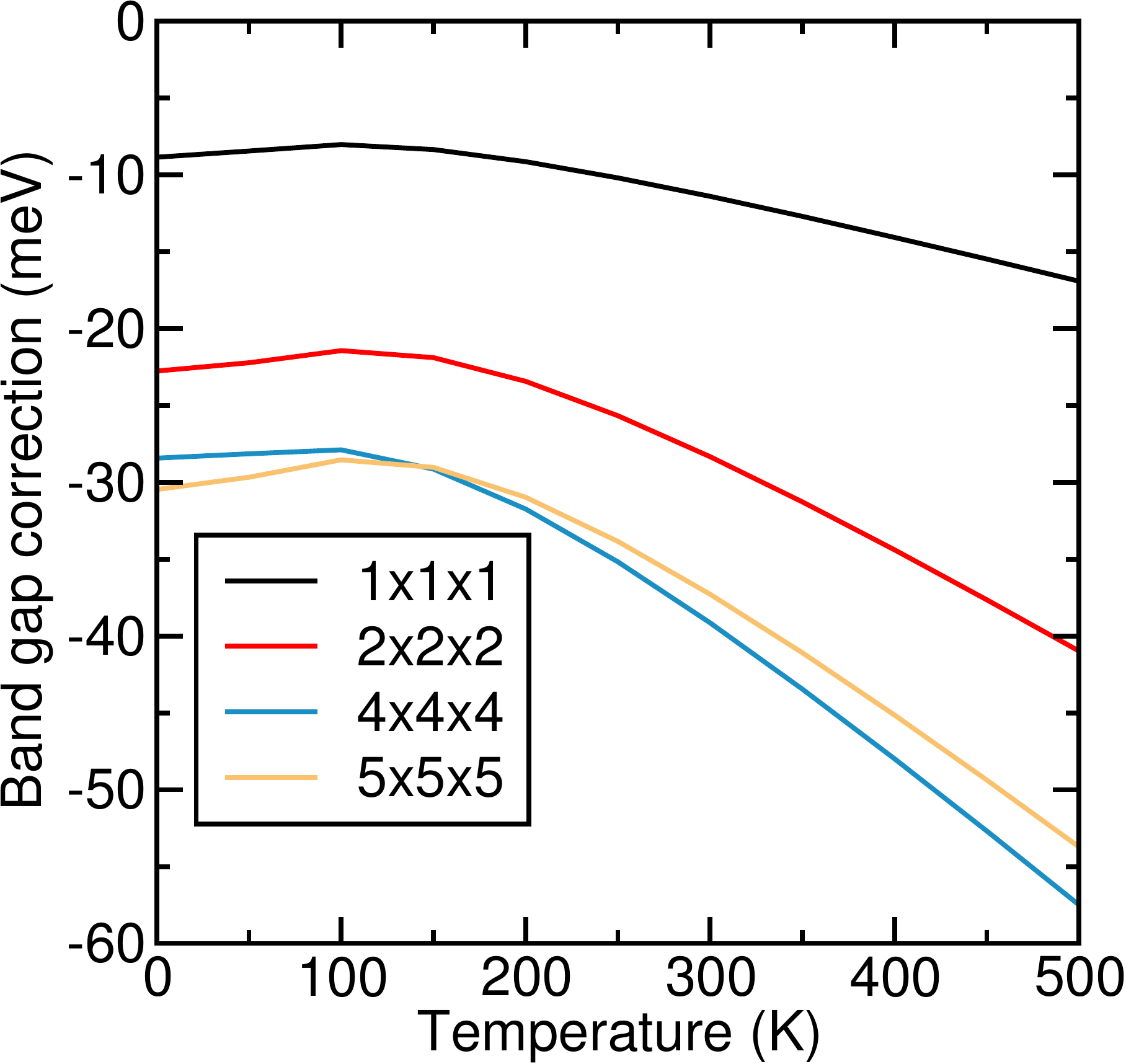}
\caption{{\footnotesize Vibrational correction (quantum and thermal) to the band gap of CZTS using the quadratic approximation and different $\mathbf{q}$-point grid sizes. The calculations have been performed using the hybrid HSE06 functional.\cite{hse03_functional,hse06_functional,hse06_functional_erratum}}}
\label{fig:qpoint}
\end{center}
\end{figure}

The electron-phonon coupling results reported in the main text correspond to a $4\times4\times4$ $\mathbf{q}$-point grid to sample the vibrational Brillouin zone. In Fig.~\ref{fig:qpoint} we report the temperature dependence of the band gap of CZTS using increasingly large $\mathbf{q}$-point grid sizes to confirm that a $4\times4\times4$ $\mathbf{q}$-point grid gives reasonably converged results when compared to the results obtained using a $5\times5\times5$ $\mathbf{q}$-point grid.

We note that to obtain the $5\times5\times5$ $\mathbf{q}$-point grid results, we fully exploited nondiagonal supercells that allow us to perform these calculations using supercells containing a maximum of $40$ atoms ($8$ atoms in the CZTS primitive cell). These calculations using the standard diagonal supercells used in most lattice dynamics codes would have required $1000$ atoms, which is computationally prohibitive using a hybrid functional.

\subsection{Mode-resolved electron-phonon coupling}

\begin{figure}
\begin{center}
\includegraphics[width=0.35\textwidth]{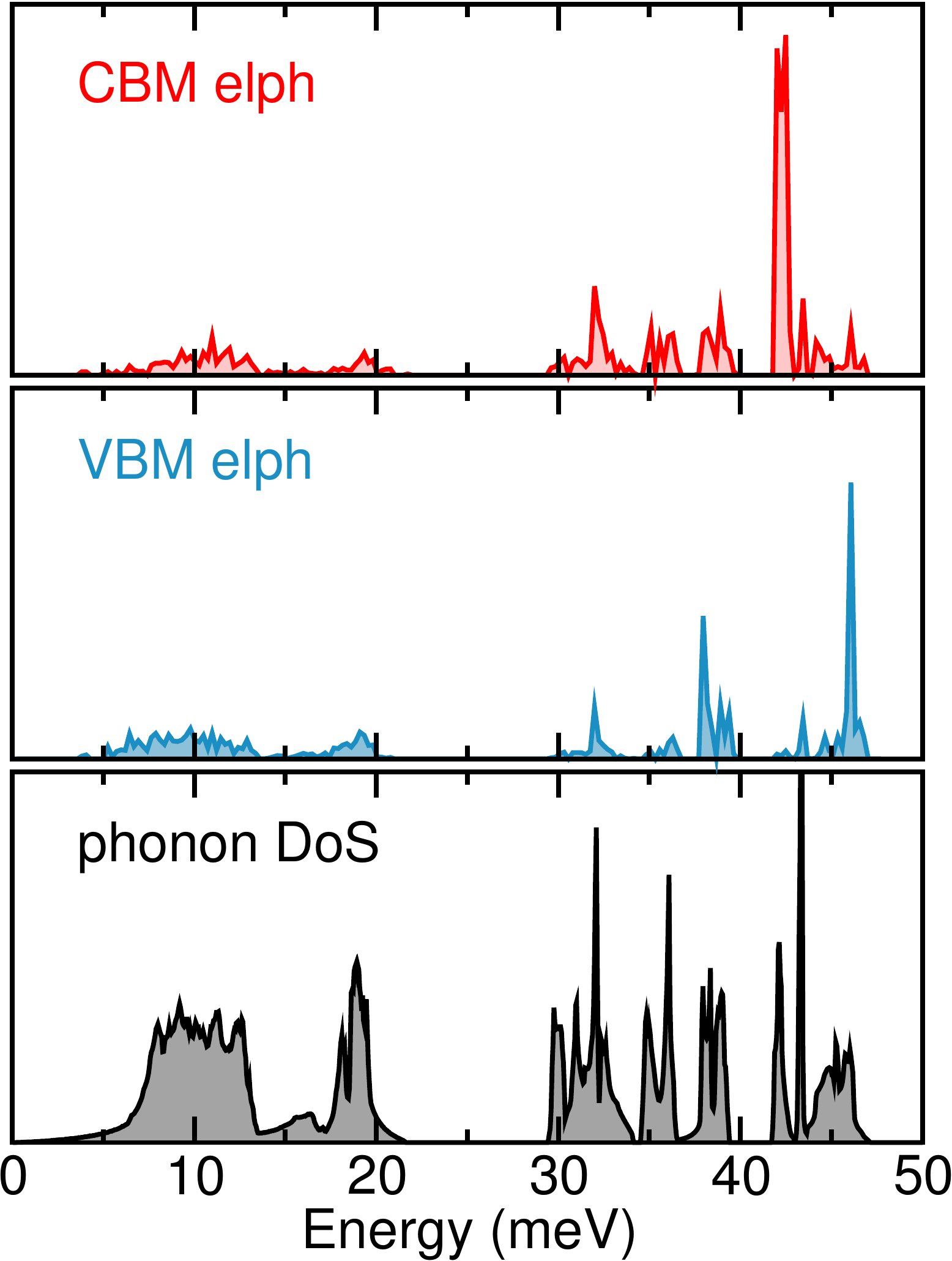}
\caption{{\footnotesize Density of states of the electron-phonon coupling to the CBM (top), to the VBM (middle), and phonon density of states (bottom) for CZTS. The calculations have been performed using the hybrid HSE06 functional.\cite{hse03_functional,hse06_functional,hse06_functional_erratum}}}
\label{fig:mode_resolved}
\end{center}
\end{figure}

We have calculated the contribution of individual phonon modes $(\mathbf{q},\nu)$ to the change in the VBM and CBM due to electron-phonon coupling arising from Eq.~(\ref{eq:quad}). The results are depicted in Fig.~\ref{fig:mode_resolved}, and demonstrate that the dominant electron-phonon coupling contributions arise from the high-energy phonon modes which correspond to vibrations dominated by the motion of the sulfur atoms.

\section{Electron-phonon coupling calculations for C\lowercase{d}S}

In this section we detail the electron-phonon coupling calculations for CdS used in the main text. Calculations have been performed with the CdS wurtzite structure.

\begin{figure}
\begin{center}
\includegraphics[width=0.35\textwidth]{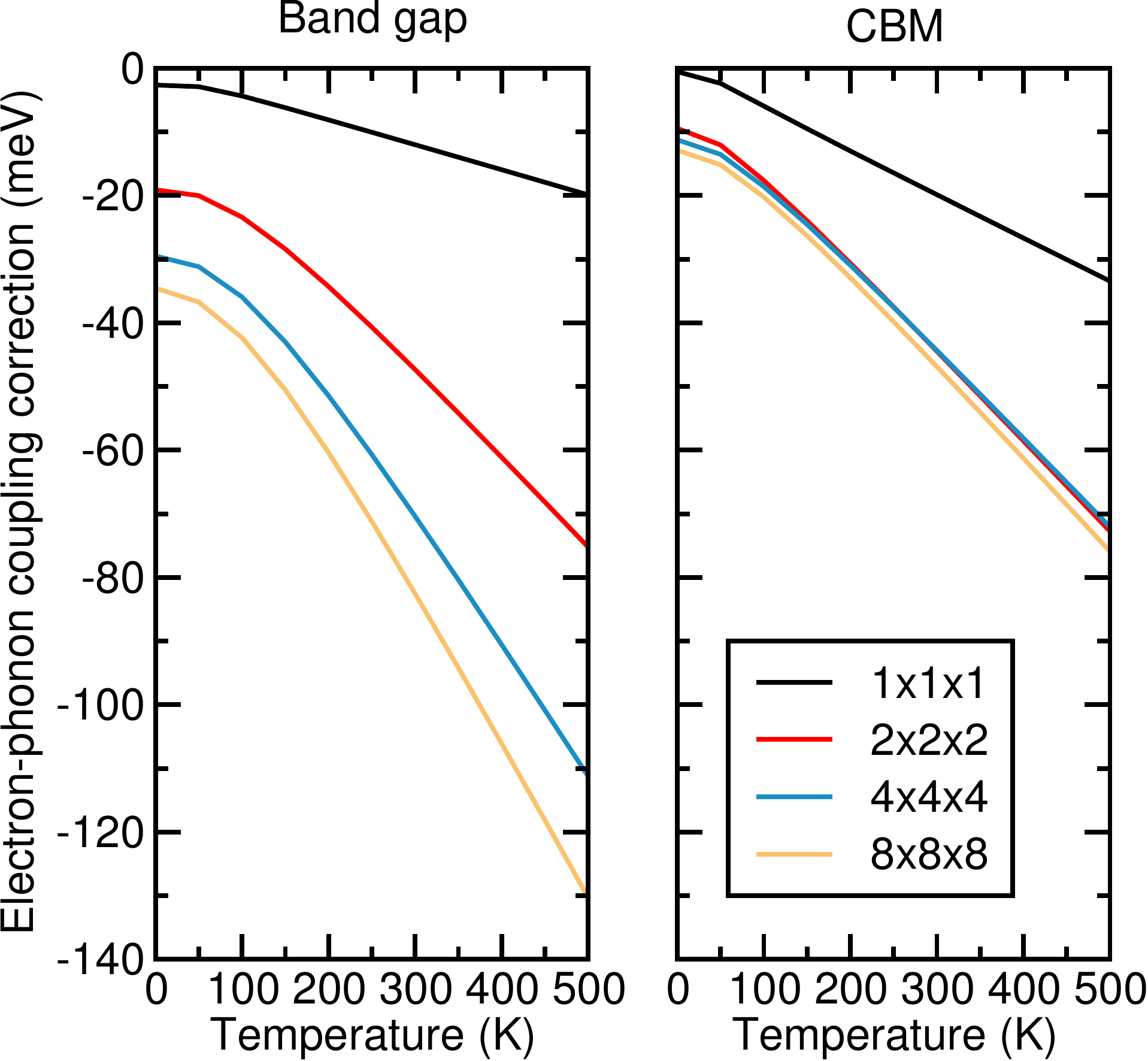}
\caption{{\footnotesize Vibrational correction (quantum and thermal) to the band gap (left) and conduction band minimum (right) of CdS. The calculations have been performed using the semilocal PBEsol functional and the quadratic approximation.}}
\label{fig:qpoint-cds}
\end{center}
\end{figure}
Using the PBEsol functional,\cite{pbesol_functional} we calculate the temperature dependence of the band gap of CdS as a function of the $\mathbf{q}$-point grid size used to sample the  vibrational Brillouin zone. We find slow convergence with respect to grid sizes up to $8\times8\times8$, as shown in Fig.~\ref{fig:qpoint-cds}. However, we note that the temperature dependence of the CBM is already well converged for $2\times2\times2$ $\mathbf{q}$-point grids, and the slow convergence of the overall band gap is determined by the slow convergence of the VBM temperature dependence. The CBM is the relevant energy eigenvalue for the band alignment discussion in the main text, and we therefore use the $2\times2\times2$ $\mathbf{q}$-point grid results. Furthermore, we note that repeating the calculations using the hybrid HSE06 functional\cite{hse03_functional,hse06_functional,hse06_functional_erratum} using a $2\times2\times2$ $\mathbf{q}$-point grid leads to a temperature dependence for the CBM in good agreement with the PBEsol results. The agreement between HSE06 and PBEsol is poor for the temperature dependence of the VBM, but again we remark that we are interested in the CBM behaviour.

\section{Thermal expansion}

\begin{figure}
\begin{center}
\includegraphics[width=0.35\textwidth]{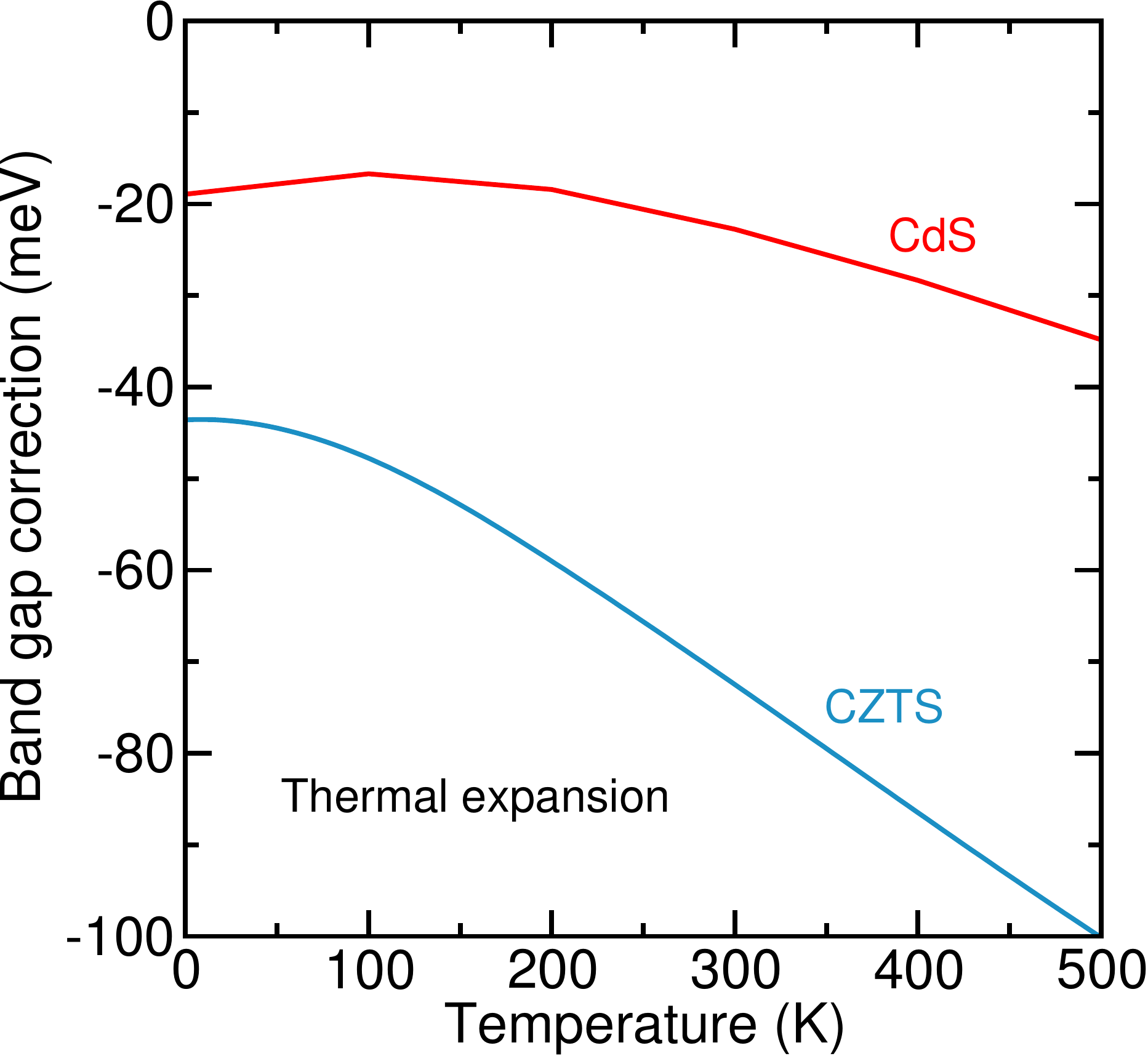}
\caption{{\footnotesize Band gap correction in CZTS and CdS arising from thermal expansion.}}
\label{fig:te}
\end{center}
\end{figure}

We have used the quasiharmonic approximation\cite{dove_lattice_dynamics_book} to calculate the thermal expansion exhibited by CZTS and CdS. We have then evaluated the change in the band gap with increasing temperature due to thermal expansion, as shown in Fig.~\ref{fig:te}. CdS exhibits a nonmonotonic temperature dependence of the band gap, with a negative quantum correction, a positive smaller curvature at small temperatures, and a turnover to a negative curvature at higher temperatures. This is a result on an analogous nonmonotonic temperature dependence in the volume of CdS.

\section{Temperature dependence of crystal field splitting of C\lowercase{u}$_2$Z\lowercase{n}S\lowercase{n}S$_4$}

\begin{figure}
\begin{center}
\includegraphics[width=0.35\textwidth]{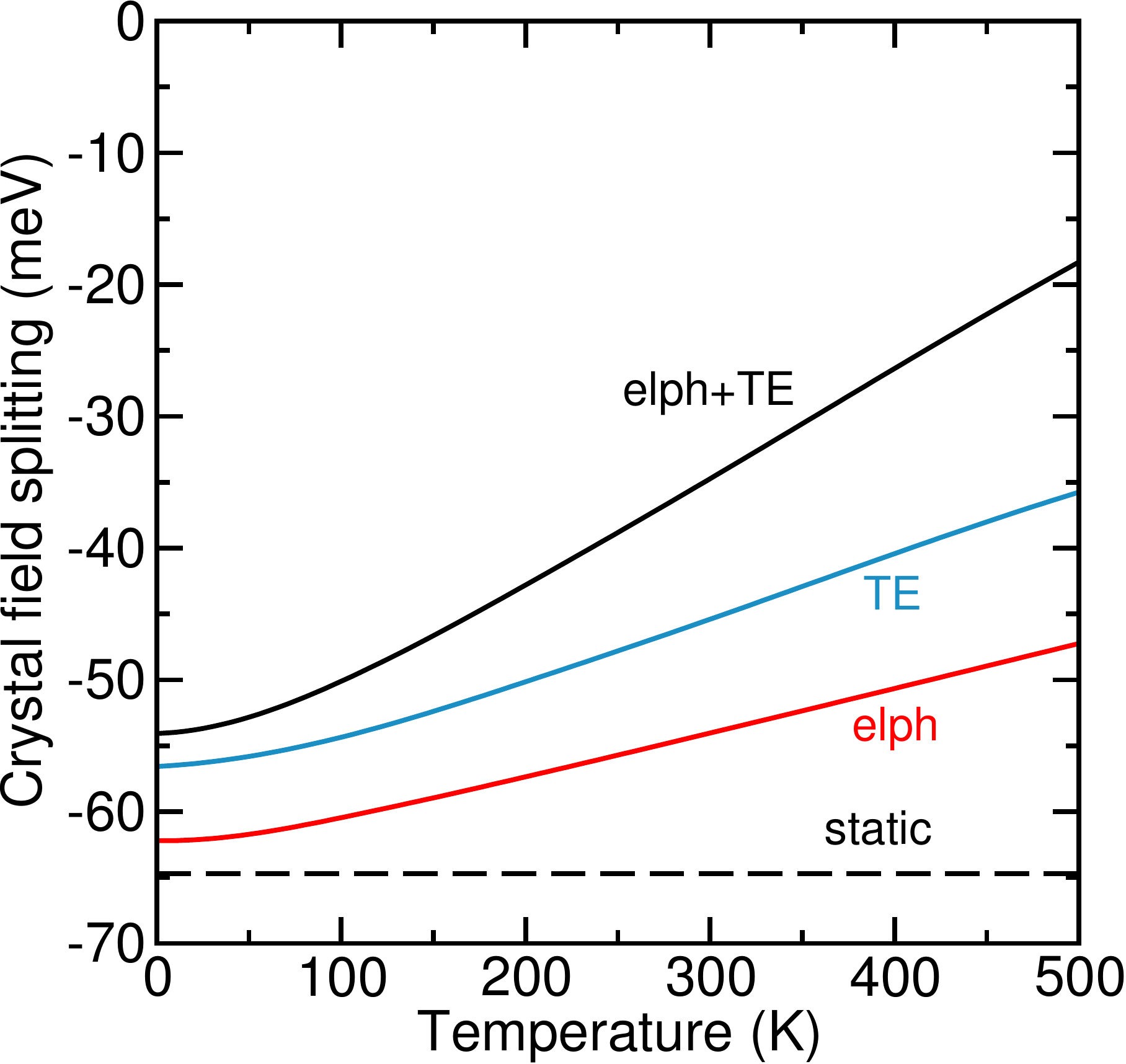}
\caption{{\footnotesize Temperature dependence of the crystal field splitting $\Gamma_{5\mathrm{v}}-\Gamma_{4\mathrm{v}}$ of the valence bands of CZTS arising from electron-phonon coupling only (elph), thermal expansion only (TE), and the total dependence (elph+TE).}}
\label{fig:cfs}
\end{center}
\end{figure}

In Fig.~\ref{fig:cfs} we show the temperature dependence of the crystal field splitting of the valence bands $\Gamma_{5\mathrm{v}}-\Gamma_{4\mathrm{v}}$ of CZTS. The calculations include the contributions of both thermal expansion and electron-phonon coupling. The crystal field splitting decreases from $-54$\,meV at $0$\,K to $-18$\,meV at $500$\,K.
\section{Vibrational density of states of C\lowercase{u}$_2$Z\lowercase{n}S\lowercase{n}S\lowercase{e}$_4$}

\begin{figure}
\begin{center}
\includegraphics[width=0.35\textwidth]{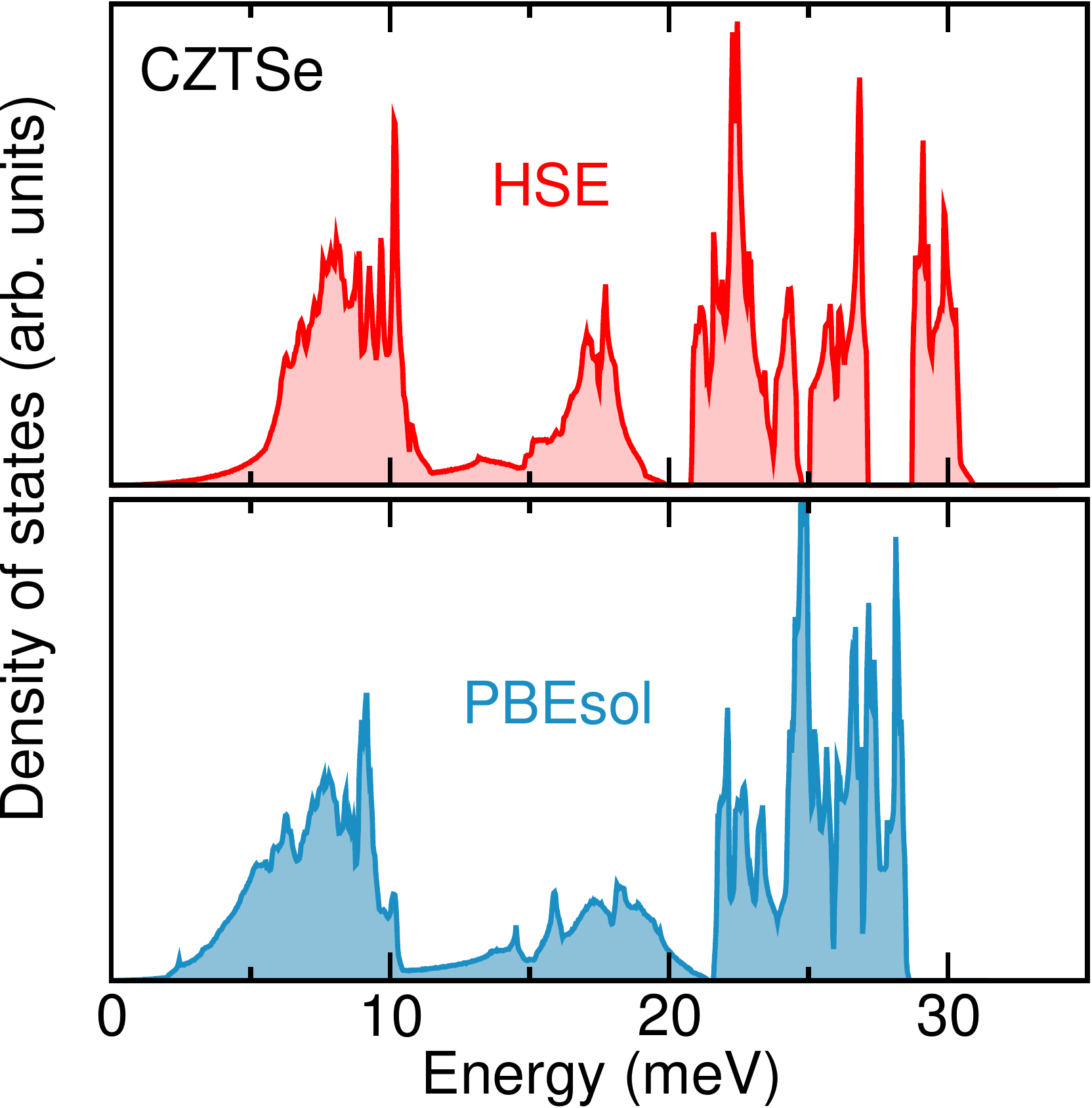}
\caption{{\footnotesize Vibrational density of states of CZTSe calculated using the HSE06 (top) and PBEsol (bottom) functionals.}}
\label{fig:cztse-dos}
\end{center}
\end{figure}

In Fig.~\ref{fig:cztse-dos} we show the vibrational density of states of Cu$_2$ZnSnSe$_4$ (CZTSe) comparing the hybrid HSE06 functional to the semilocal PBEsol functional. Similarly to the CZTS results reported in the main text, we find that the HSE06 high-energy frequencies are larger than the corresponding PBEsol frequencies.

\end{document}